\documentclass[amssymbols,11pt]{article}
\newcommand{\curly}[2]
{\left( \begin{array}{c} #1 \\ #2 \end{array} \right)}

\newcommand{\curl}[4]
{\left( \begin{array}{cc} #1 & #2 \\[2ex] #3 & #4 \end{array} \right)}

\newcommand\inta{\int^{\infty}_{-\infty}}
\newcommand\into{\int^{\infty}_{0}}

\newcommand\half{ \frac{1}{2} }

\newcommand\s{Schr\"odinger }

\renewcommand\Im{ {\mathrm{Im}}}

\input epsf

\usepackage{setspace}

\begin{document}  
\title{The Davey-Stewartson I Equation on the Quarter \\
Plane with Homogeneous Dirichlet Boundary Conditions}
\author{A.S. Fokas \\
{\em Department of Applied Mathematics and Theoretical Physics} \\
{\em University of Cambridge} \\
{\em Cambridge, CB3 0WA, UK} \\
{\em t.fokas@damtp.cam.ac.uk}}
\date{May 2003}

\maketitle

\doublespacing

\centerline{\bf Abstract}
Dromions are exponentially localised coherent structures supported by
nonlinear integrable evolution equations in two spatial dimensions. In
the study of \emph{initial-value} problems on the plane, such
solutions occur only if one imposes nontrivial boundary conditions at
infinity, a situation of dubious physical significance. However it is
established here that dromions appear naturally in the study of
\emph{boundary-value} problems.  In particular, it is shown that the
long time asymptotics of the solution of the Davey-Stewartson I
equation in the quarter plane with arbitrary initial conditions and
with zero Dirichlet boundary conditions is dominated by dromions. The
case of non-zero Dirichlet boundary conditions is also discussed. 

\pagebreak

\section{Introduction}
We consider the Davey-Stewartson (DS) I equation [1] 
$$ iq_t + \half (q_{xx} + q_{yy}) - (\varphi_x + |q|^2)q=0,$$
$$ \varphi_{xx} - \varphi_{yy} + 2|q|^2x=0. \eqno (1.1)$$
In the context of water waves this equation is the shallow water limit of the
Benney-Roskes [2] equation in the case of dominant surface tension.  In this
case $q(x,y,t)$ is the amplitude of a surface wave packet and $\varphi$ is the
velocity potential of the associated mean flow.  In general the DSI equation
provides a two-dimensional generalization of the nonlinear \s equation and can be
derived from general asymptotic considerations [3].

We introduce characteristic coordinates and we also replace the second order
equation (1.1b) by two first order equations: Let $\xi,\eta$, $U_1(\xi,\eta,t)$,
$U_2(\xi,\eta,t)$, be defined by
$$ \xi = x+y, \quad \eta =x-y, \quad U_1 = -\varphi_\eta - \half |q|^2, \quad
U_2 =- \varphi_\xi - \half |q|^2. \eqno (1.2)$$
Then the DSI equation becomes
$$ iq_t + q_{\xi\xi} + q_{\eta\eta} + (U_1+U_2)q=0,$$
$$U_{1_\xi} = \half |q|^2_\eta, \quad U_{2_\eta} = \half |q|^2_\xi. \eqno (1.3)$$
Indeed, writing equation(1.1a) in characteristic coordinates and using the
definitions of $U_1,U_2$, we find equation (1.3a).  Also the definition of $U_1$
implies $U_{1_\xi} =- \varphi_{\eta\xi} - |q|^2_\xi/2$; writing equation (1.1b) in
characteristic coordinates and using this equation to replace $\varphi_{\eta\xi}$,
we find equation (1.3b).  Similarly for equation (1.3c).

\paragraph{Formulation of the Problem.} 

Let the complex-valued function $q(\xi,\eta,t)$ and the real-valued functions
$U_1(\xi,\eta,t)$, $U_2(\xi,\eta,t)$, satisfy equations (1.3) in the domain
$$ 0<\xi<\infty, \quad 0<\eta<\infty, \quad t>0, \eqno (1.4)$$
with the following initial and boundary conditions:
$$ q(\xi,\eta,0) = q_0(\xi,\eta), \quad 0<\xi<\infty, \quad 0<\eta<\infty,$$
$$q(0,\eta,t) = 0, \quad U_1(0,\eta,t) = u_1(\eta,t), \quad 0<\eta<\infty, \quad
t>0, $$
$$ q(\xi,0,t) =0, \quad U_2(\xi,0,t) = u_2(\xi,t), \quad 0<\xi<\infty, \quad t>0,
\eqno (1.5)$$
where the functions $q_0(\xi,\eta)$, $u_1(\eta,t)$, $u_2(\xi,t)$ have sufficient
smoothness and they also decay as $\xi\rightarrow \infty$ and
$\eta\rightarrow\infty$.

\paragraph{Notations.}

\begin{itemize}
\item
Bar denotes complex conjugation.

\item
$M_{11},M_{12},M_{21},M_{22}$ denote the $(11),(12),(21),(22)$ entries
  of the $2\times2$ matrix $M$.

\item
$M_{D},M_{0}$ denote the diagonal and the off-diagonal parts of the
  $2\times2$ matrix $M$.
\end{itemize}

\paragraph{Theorem 1.1}  Given $q_0(\xi,\eta)$ define the vector
$(M_1(\xi,\eta,k), M_2(\xi,\eta,k))^T$ by 
$$ M_{1_\xi} -ikM_1 = -\half q_0M_2, \quad M_{2_\eta} = \half \bar q_0M_1, \quad
0<\xi<\infty, \quad 0<\eta<\infty, \quad \Im{k} \leq 0, $$
$$ \lim_{\xi\rightarrow\infty} M_1 =0, \quad M_2(\xi,0,k) = 1. \eqno (1.6)$$
Given $q_0$ and $M_2$ define $S_0(k,l)$ by
$$ S_0(k,l) = \frac{1}{4\pi} \into \into d\xi d\eta q_0(\xi,\eta) M_2(\xi,\eta,k)
e^{-ik\xi-il\eta}, \quad \Im k \leq 0, \quad \Im l \leq 0. \eqno (1.7)$$
Given $S_0$, $u_1$, $u_2$, define $\hat S(\xi,\eta, t)$ by
$$ i\hat S_t + \hat S_{\xi\xi} + \hat S_{\eta\eta} + (u_1(\eta,t) +
u_2(\xi,t))\hat S =0, \quad 0<\xi < \infty, \quad 0<\eta<\infty, \quad t>0,$$
$$ \hat S(\xi,\eta,0) = \inta \inta dkdl e^{ik\xi + il\eta} S_0(k,l), \quad
0<\xi<\infty, \quad 0<\eta<\infty, $$
$$ \hat S(\xi,0,t) = 0, \quad 0<\xi<\infty, \quad t>0,$$
$$ \hat S(0,\eta,t) =0, \quad 0<\eta <\infty, \quad t>0. \eqno (1.8)$$
Given $\hat S$ define the $2\times 2$ matrices $M^+(\xi,\eta,t,k)$ and
$M^-(\xi,\eta,t,k)$ as the solution of the following non-local Riemann-Hilbert
problem:

\begin{itemize}
\item $M^+$ and $M^-$ are analytic for $\Im k>0$ and $\Im k<0$ respectively.
\item $M^\pm = I + O(\frac{1}{k})$, $k\rightarrow \infty, \quad \Im k
  \neq 0$.
\item For real $k$, $M^+$ and $M^-$ satisfy the following jump condition,
\end{itemize}
$$\curly{M^+_{11}}{M^+_{21}} (\xi,\eta,t,k) - \curly{M^-_{11}}{M^-_{21}}
(\xi,\eta,t,k) = - \inta dl \overline{S(l,k,t)} e^{-il\xi-ik\eta}
\curly{M^+_{12}}{M^+_{22}} (\xi,\eta,t,l), $$
$$ \curly{M^+_{12}}{M^+_{22}} (\xi,\eta,t,k) - \curly{M^-_{12}}{M^-_{22}}
(\xi,\eta,t,k) = -\inta dl S(k,l,t)e^{il\eta + ik\xi} \curly{M^-_{11}}{M^-_{21}}
(\xi,\eta, t,l), \eqno (1.9)$$
where
$$ S(k,l,t) = \frac{1}{4\pi^2} \into \into d\xi d\eta \hat S(\xi,\eta,t)e^{-
ik\xi-il\eta}. \eqno (1.10)$$
This Riemann-Hilbert problem has a unique global solution.  

Define $q(\xi,\eta,t)$ by
$$ q = 2i\lim_{k\rightarrow\infty}(kM_{12}). \eqno (1.11)$$
Then $q$ satisfies (1.3), (1.5).

\paragraph{Remark 1.1}  Although the evolution of the scattering data $S(k,l,t)$
is in principle determined by equations (1.8), the relevant time dependence is
complicated.  In turn, this makes it difficult to determine the long time behavior
of the solution $M$ of the RH problem (1.9).  This difficulty can be bypassed by
formulating an {\it inverse problem} for $S(k,l,t)$.  Since equation
(1.8a) is precisely the one studied in [4], the relevant analysis is
identical with the one presented in [4]: (a) If $u_1$ and $u_2$ are time-independent
then the analysis of (1.8) is intimately related with the analysis of the
time-independent \s equation
$$\psi_{xx} + (u(x) +k^2) \psi =0, $$
where $u$ is either $u_1$ or $u_2$.  After  a long 
time the solution of (1.8) is dominated by the {\it discrete} spectrum of $u_1$ and of
$u_2$.  If $u_1$ and $u_2$ have $N_1$ and $N_2$ discrete eigenvalues respectively then
the long time asymptotics of $q$ is given by an $(N_1,N_2)$-\emph{breather}
solution.  (b) If $u_1$ and $u_2$ are time-dependent then the analysis
of (1.8) is related with the analysis of the time-dependent \s
equation.  Assuming a certain completeness relation for the
eigenfunctions of the time-dependent \s equation, the long time
asymptotics of the solution $q$ is dominated again by the associated
discrete spectrum; in this case the solution is
dominated by an $(N_1,N_2)$-\emph{dromion} solution.

\paragraph{Remark 1.2}
The explicit form of the $(N_1,N_2)$-dromion solution can be found in
[4], see also [5],[6].  Dromion solutions are exponentially decaying in
both $\xi$ and $\eta$. However, in contrast to the one-dimensional
solitons, these solutions do \emph{not} preserve their form upon
interaction and therefore can exchange energy. These coherent
strucures can be driven everywhere in the quarter-plane $\xi \geq 0, \eta
\geq 0$, by choosing a suitable motion for $u_1(\eta,t), u_2(\xi,t)$.

\paragraph{Remark 1.3}
A characteristic feature of boundary-value problems for integrable
evolution PDEs in one spatial dimension is that the (spectral)
functions defining the associated Riemann-Hilbert problems \emph{cannot}
in general be expressed explicitly in terms of the given boundary
conditions [7].  Explicit formulae can be obtained only for a
particular class of boundary conditions which are  referred to in [7]
as \emph{linearisable}.  The situation is similar for boundary-value
problems in two spatial dimensions.  We emphasize that equations (1.8)
are explicitly defined in terms of the given initial conditions, thus the
homogeneous Dirichlet problem (1.3)--(1.5) belongs to the class of
linearisable conditions.  It is shown below that for non-homogeneous
Dirichlet boundary conditions, the equations defining $\hat{S}$, in addition
to the \emph{given} Dirichlet boundary conditions, also involve the
\emph{unknown} Neumann boundary values.  

\paragraph{Proposition 1.1} Let the complex-valued function $q(\xi,\eta,t)$
satisfy a non-homogeneous Dirichlet boundary-value problem in the
quarter plane, namely equations (1.3)--(1.5), where $q(0,\eta,t)=0$
and $q(\xi,0,t)=0$ are replaced by
$$
q(0,\eta,t) = g_{0}(\eta,t), \quad q(\xi,0,t) = f_{0}(\xi,t), \eqno{(1.12)}
$$
and the functions $g_0, f_0$ have sufficient smoothness and they also decay
as $\eta \to \infty, \xi \to \infty$.  \emph{Assume} that there
exists a global solution.

This solution can be expressed by equation (1.11) through the solution
of the Riemann-Hilbert problem defined in Theorem 1.1.  This problem
is uniquely defined in terms of the function $\hat{S}(\xi,\eta,t)$
which satisfy the following:
\begin{itemize}
\item $\hat{S}$ solves the linear evolution PDE
$$
i\hat{S}_{t} + \hat{S}_{\xi\xi} + \hat{S}_{\eta\eta} +
(u_{1}+u_{2})\hat{S} + \int_{0}^{\xi} d\tilde{\xi}
F_{1}(\xi,\tilde{\xi},t)\hat{S}(\tilde{\xi},\eta,t) + \int_{0}^{\eta}
d\tilde{\eta} F_{2}(\eta,\tilde{\eta},t)\hat{S}(\xi,\tilde{\eta},t) =
0, \eqno (1.13)
$$
where
$$\begin{array}{l}
F_{1}(\xi,\tilde{\xi},t) = \frac{1}{4} [
  \bar{f}_{1}(\tilde{\xi},t) f_{0}(\xi,t) -
  \bar{f}_{0}(\tilde{\xi},t) f_{1}(\xi,t)] - \frac{1}{16}
  f(\xi,t) \bar{f}(\tilde{\xi},t) \int_{\tilde{\xi}}^{\xi} d\xi'
  |f(\xi',t)|^{2}, \\
F_{2}(\eta,\tilde{\eta},t) = \frac{1}{4} [
  \bar{g}_{1}(\tilde{\eta},t) g_{0}(\eta,t) -
  \bar{g}_{0}(\tilde{\eta},t) g_{1}(\eta,t)] - \frac{1}{16}
  g(\eta,t) \bar{g}(\tilde{\eta},t) \int_{\tilde{\eta}}^{\eta} d\eta'
  |g(\eta',t)|^{2},
\end{array}$$
and $g_1(\eta,t), f_1(\xi,t)$ denote the Neumann boundary values
  $q_{\xi}(0,\eta,t), q_{\eta}(\xi,0,t)$;
\item $\hat{S}$ satisfies the initial and boundary conditions
$$
\hat{S}(\xi,\eta,0) = \hat{S}_{0}(\xi,\eta), \quad \hat{S}(\xi,0,t) =
\pi f_{0}(\xi,t), \quad \hat{S}(0,\eta,t) = \pi g_{0}(\eta,t),
\eqno{(1.14)}
$$
where $\hat{S}_{0}(\xi,\eta)$ is defined in terms of $q_{0}(\xi,\eta)$
by the rhs of equation (1.7).
\end{itemize}

\paragraph{Remark 1.4} 
In the case of boundary-value problems for evolution PDEs
in one spatial dimension, the unknown boundary values can be
characterised in terms of the given boundary conditions through the
analysis of certain \emph{global relations} [7],[8]. Such global
relations exist in two spatial dimensions and can also be used for the
characterisation of the unknown boundary values.  This analysis, which
is rather complicated, will be presented elsewhere. 

\paragraph{Organisation of the paper.}
The DSI equation admits a Lax pair formulation. The $t$-independent part
of the Lax pair (see equations (4.2)) is analysed in section 2. The
specific form of the $t$-part of the Lax pair (see equations (3.4))
depends on the specific form of the boundary conditions.  The $t$-part of
the Lax pair for the homogeneous and non-homogeneous Dirichlet cases
is discussed in sections 3 and 5 respectively.  The proof of Theorem
1.1 is presented in section 4.  Section 6 contains further
discussion.

\section{The $t$-independent part of the Lax Pair}

Throughout this section we \emph{suppress the $t$-dependence}.  The
$t$-independent part of the Lax pair is given by equation (4.2).
Analysing this equation in characteristic coordinates we find the
following:
\paragraph{Proposition 2.1} Let the vectors
$$\curly{M^+_{11}}{M^+_{21}}, \curly{M^+_{12}}{M^+_{22}}, \curly{M^-_{11}}{M^-_{21}},
\curly{M^-_{12}}{M^-_{22}}, \eqno (2.1)$$
which are functions of $\xi,\eta, k$, be defined by
$$\begin{array}{ll}
M^+_{11} = 1 - \half \int^\xi_0 d\xi' qM^+_{21}, & M^+_{12} = - \half \int^\xi_0 d\xi'
e^{ik(\xi-\xi')} qM^+_{22}, \\ \\
M^+_{21} = -\half \int^\infty_\eta d\eta' e^{ik(\eta'-\eta)} \bar qM^+_{11}, &
M^+_{22} = 1+\half \int^\eta_0 d\eta' \bar qM^+_{12}, \\ \\
M^-_{11} = 1-\half \int^\xi_0 d\xi' qM^-_{21}, & M^-_{12} = \half \int^\infty_\xi
d\xi'e^{ik(\xi-\xi')} qM^-_{22}, \\ \\
M^-_{21} = \half \int^\eta_0 d\eta' \bar qe^{ik(\eta'-\eta)}M^-_{11}, & M^-_{22} = 1 +
\half \int^\eta_0 d\eta' \bar qM^-_{12}, \end{array} \eqno (2.2)$$
where $q(\xi,\eta)$ has sufficient smoothness and decay. Then:

1. The first two vectors in (2.1) are analytic in $k$ for $\Im k>0$, while the last
two vectors in (2.1) are analytic for $\Im k <0$.

2. For real $k$, the vectors (2.1) satisfy the relationships

$$\curly{M^+_{11}}{M^+_{21}}(\xi,\eta,k) -  \curly{M^-_{11}}{M^-_{21}}(\xi,\eta,k) = -
\inta dl \overline{S(l,k)} e^{-il\xi-ik\eta} \curly{M^+_{12}}{M^+_{22}} (\xi,\eta,l),
$$

$$\curly{M^+_{12}}{M^+_{22}}(\xi,\eta,k) -  \curly{M^-_{12}}{M^-_{22}}(\xi,\eta,k) = -
\inta dl  S(k,l)  e^{il\eta+ik\xi} \curly{M^-_{11}}{M^-_{21}} (\xi,\eta,l),
\eqno (2.3)$$
where $S(k,l)$ is defined by
$$ S(k,l) = \frac{1}{4\pi} \into \into d\xi d\eta q(\xi,\eta) M^-_{22}(\xi,\eta,k)
e^{-ik\xi-il\eta}, \quad \Im k \leq 0, \quad \Im l \leq 0. \eqno (2.4)$$

3. The vectors (2.1) have the following behaviour for large $k$,
$$M^\pm_{11} = 1+O\left(\frac{1}{k}\right), \quad M^\pm_{22} = 1 +O\left(\frac{1}{k}\right), \quad
M^\pm_{12} = O\left(\frac{1}{k}\right), \quad M^\pm_{21} = O\left(\frac{1}{k}\right), \quad k\rightarrow
\infty, \quad \Im k \neq 0. \eqno (2.5)$$
\paragraph{Proof} \ \ 

1. The vector $(M^+_{11}, M^+_{21})^{T}$ satisfies a system of linear Volterra integral
equations with kernel analytic in $k$ for $\Im k>0$ (since $\eta' \geq \eta)$.  Thus
this vector is analytic for $\Im k>0$.  Similarly for the other vectors.

2. Let the matrices $\Psi^+(\xi,\eta,k)$ and $\Psi^-(\xi,\eta,k)$ be defined by
$$ \Psi^\pm =
\curl{M^\pm_{11}e^{ik\eta}}{M^\pm_{12}e^{-ik\xi}}{M^\pm_{21}e^{ik\eta}}{M^\pm_{22}e^{-
ik\xi}}. \eqno (2.6)^\pm $$
Then 
$$\begin{array}{ll}
\Psi^+_{11} = e^{ik\eta} - \half \int^\xi_0 d\xi'q\Psi^+_{21}, & \Psi^+_{12} = -\half
\int^\xi_0d\xi'q\Psi^+_{22}, \\ \\ 
\Psi^+_{21} = \half \int^\infty_\eta d\eta' \bar q\Psi^+_{11}, & \Psi^+_{22} =
e^{-ik\xi} + \half \int^\eta_0 d\eta' \bar q \Psi^+_{12}, \\ \\
\Psi^-_{11} = e^{ik\eta} - \half \int^\xi_0 d\xi'q\Psi^-_{21}, & \Psi^-_{12} = \half
\int^\infty_\xi d\xi' q \Psi^-_{22}, \\ \\
\Psi^-_{21} = \half \int^\eta_0 d\eta' \bar q \Psi^-_{11}, & \Psi^-_{22} = e^{-ik\xi}
+ \half \int^\eta_0 d\eta' \bar q\Psi^-_{12}. \end{array} \eqno (2.7)$$
Subtracting the equations defining the vectors $(\Psi^+_{12}, \Psi^+_{22})^T$ and
$(\Psi^-_{12}, \Psi^-_{22})^T$ we find
$$ \Psi^+_{12} - \Psi^-_{12} = -\half \into d\xi' q \Psi^-_{22} - \half \int^\xi_0
d\xi' q(\Psi^+_{22} - \Psi^-_{22}), $$
$$\Psi^+_{22} - \Psi^-_{22} = \half \int^\eta_0 d\eta' \bar q(\Psi^+_{12} -
\Psi^-_{12}). \eqno (2.8)$$
Using the definition of $S(k,l)$ it follows that the first term of the rhs of
equation (2.8a) equals
$$- \inta dl S(k,l)e^{il\eta}.$$
Comparing equations (2.8) with the equation satisfied by $(\Psi^-_{11}, \Psi^-_{21})^T$
we find equation (2.3b) written in terms of the functions $\Psi^\pm$ instead of the
functions $M^\pm$. 

Similarly subtracting the equations defining the vectors $(\Psi^+_{11}, \Psi^+_{21})^T$
and $(\Psi^-_{11}, \Psi^-_{21})^T$ we find equation (2.3a), where instead of
$-\overline{S(l,k)}$ we have the function
$$T(k,l) = -\frac{1}{4\pi}  \into \into d\xi d\eta \bar
q(\xi,\eta)M^+_{11}(\xi,\eta,k)e^{ik\eta + il\xi}. \eqno (2.9)$$
We will now show that
$$T(k,l) =- \overline{S(l,k)}. \eqno (2.10)$$
Indeed, the vectors $(\Psi^+_{11},\Psi^+_{21})^T$ and $(\Psi^-_{12}, \Psi^-_{22})^T$
satisfy the equations
$$ \Psi^+_{11_\xi} = - \half q\Psi^+_{21}, \qquad \overline{\Psi^-_{12_\xi}} = - \half
\bar q \overline{\Psi^-_{22}} $$

$$ \Psi^+_{21_\xi} =   \half \bar q\Psi^+_{11}, \qquad \overline{\Psi^-_{22_\eta}}
 =   \half
q \overline{\Psi^-_{12}}.$$
Hence 
$$\left( \Psi^+_{11}(k)  \overline{\Psi^-_{12}(l)} \right)_\xi = - \left( \Psi^+_{21}(k) 
\overline{\Psi^-_{22}(l)}\right)_\eta, $$
where for convenience of notation we have suppressed the $(\xi,\eta)$ dependence.
Integrating this equation we find
$$\into d\eta \left[ \left. \Psi^+_{11}(k)\overline{\Psi^-_{12}(l)}\right|_{\xi=\infty}
-\left. \Psi^+_{11}(k)\overline{\Psi^-_{12}(l)}\right|_{\xi=0}\right] =$$
$$ = -\into d\xi \left[ \left. \Psi^+_{21}(k)\overline{\Psi^-_{22}(l)}\right|_{\eta=0}
- \left. \Psi^+_{21}(k)\overline{\Psi^-_{22}(l)}\right|_{\eta=0}\right].$$
Using equations (2.7) to compute the boundary values appearing above, for example
$$ \left. \Psi^-_{12}\right|_{\xi=\infty} = \left. \Psi^+_{21}\right|_{\eta=\infty}
=0, \quad
\left. \Psi^+_{11}(k)\right|_{\xi=0}=e^{ik\eta}, \quad \left.
\Psi^-_{22}(l)\right|_{\eta=0} = e^{-il\xi},$$
we find
$$ \into d\eta e^{ik\eta} \into d\xi' \bar q \overline{\Psi^-_{22}(l)} = \into d\xi
e^{il\xi} \into d\eta' \bar q \Psi^+_{11}(k).$$
The lhs of this equation is the complex conjugate of $S(l,k)$, while the rhs equals
$-T(k,l)$.

3. Equations (2.2) and integration by parts, imply equations (2.5).
\begin{flushright}
\textbf{QED}
\end{flushright}

\section{The t-part of the Lax Pair for the homogenous Dirichlet case}

In what follows we first derive the $t$-part of the Lax pair for the
vector $(\Psi^{-}_{12},\Psi^{-}_{22})^{T}$.

\paragraph{Proposition 3.1} Assume that there exists a function
$q(\xi,\eta,t)$ with sufficient smoothness and decay which satisfies
equations (1.3)--(1.5).  Let the vector $\Psi = (\Psi_1,\Psi_2)^T$ satisfy
$$\Psi_1(\xi,\eta,t,k) = \half \int^\infty_\xi
d\xi'q(\xi',\eta,t)\Psi_2(\xi',\eta,t,k),$$
$$\Psi_2(\xi,\eta,t,k) = e^{-ik\xi} + \half \int^\eta_0d\eta' \bar q(\xi,\eta',t)
\Psi_1(\xi,\eta',t,k), \quad \Im k \leq 0. \eqno (3.1)$$
Then the function $\psi_1(\eta,t,k)$ defined by
$$ \psi_1(\eta,t,k) =\Psi_1(0,\eta,t,k),$$
solves
$$ i\psi_{1_t} + \psi_{1_{\eta\eta}} + u_1(\eta,t)\psi_1 - (k^2\psi_1 + \half
q_\xi(0,\eta,t)) - i\inta dl\gamma(k-l,t)\psi_1(\eta,t,l) =0, \eqno (3.2)$$
where
$$\gamma(k,t) = \frac{i}{2\pi} \into d\xi u_2(\xi,t)e^{-ik\xi}. \eqno (3.3)$$
\paragraph{Proof}  We will first show that if $(\Psi_1,\Psi_2)^T$ satisfies equations
(3.1) and $q(\xi,\eta,t)$ satisfies equations (1.3)--(1.5),then,
$$ \begin{array}{ll}
\Psi_{1_t} = i(\partial_\xi - \partial_\eta)^2\Psi_1 +
iq(\partial_\xi-\partial_\eta)\Psi_2 + iU_1\Psi_1 -iq_\eta\Psi_2 + v_1, \\ \\
\Psi_{2_t} =-i(\partial_\xi - \partial_\eta)^2\Psi_2 -
i\bar q(\partial_\xi-\partial_\eta)\Psi_1 - iU_2\Psi_2 -i\bar q_\xi\Psi_1 + v_2,
\end{array} \eqno (3.4)$$
where the vector $v=(v_1,v_2)^T$ is given by
$$ v(\xi,\eta,t,k) = -ik^2\Psi(\xi,\eta,t,k) + \inta
dl\gamma(k-l,t)\Psi(\xi,\eta,t,l). \eqno (3.5)$$
Indeed, the vector $\Psi = (\Psi_1,\Psi_2)^T$ satisfies
$$ \Psi_{1_\xi} = -\half q\Psi_2, \quad \Psi_{2_\eta} = \half \bar q \Psi_1. \eqno
(3.6)$$
Suppose that $\Psi$ satisfies equations similar with (3.4) but with $v=0$.  It is
straightforward to verify that the compatibility condition of equations (3.4) with
$v=0$, and of equations (3.6) yields equations (1.3).  Actually, if equations (1.3) are
valid then equations (3.4) and (3.6) are compatible for any vector $v$
given by
$$v(\xi,\eta,t,k) = \inta dl\Gamma (k,l,t)\Psi(\xi,\eta,t,l).$$
The precise form of $\Gamma$ depends on the boundary conditions that $\Psi$ and $q$
satisfy.  Indeed, $\Psi_1$ can be writing in the form 
$$ \Psi_1 =-\int^\infty_\xi d\xi'\Psi_{1_{\xi'}}, \ \ {\mathrm{or}} \ \ \Psi_{1_t} =
-\int^\infty_\xi d\xi' (\Psi_{1_t})_{\xi'}. $$
By replacing in the above equation $\Psi_{1_t}$ by $A+v_1$, where $A$ is defined in
the rhs of equation (3.4a), we find
$$v_1 = -\lim_{\xi\rightarrow\infty} \left[ i(\partial_\xi-\partial_\eta)^2 \Psi_1 +
iq(\partial_\xi-\partial\eta) \Psi_2 + iU_1\Psi_1-iq_\eta\Psi_2\right] -
\int^\infty_\xi d\xi'v_{1_{\xi'}}.$$
Using
$$\lim_{\xi\rightarrow\infty} q = \lim_{\xi\rightarrow\infty} \Psi_1 =0,$$
the term in the bracket vanishes.  Also 
$$v_{1_\xi} = \inta dl \Gamma \Psi_{1_\xi} = \inta dl \Gamma \left( -
\frac{q_1}{2}\Psi_2\right) = - \frac{q_1}{2}v_1.$$
Hence
$$v_1 = \half \int^{\infty}_{\xi} d\xi'q_1v_2. \eqno (3.7a)$$
Similarly starting with
$$ \Psi_2 = \int^\eta_0 d\eta' \Psi_{2_{\eta'}} + e^{-ik\xi} \ \ {\mathrm{or}} \ \
\Psi_{2_t} = \int^\eta_0 d\eta'(\Psi_{2_t})_{\eta'}$$
we find
$$v_2 = \left[ i(\partial_\xi-\partial_\eta)^2 \Psi_2 + i\bar
q(\partial_\xi-\partial_\eta)\Psi_1 + iU_2\Psi_2 + i\bar q_\xi \Psi_1\right]_{\eta =0}
+ \half \int^\eta_0 d\eta' \bar q v_1.$$
Using
$$ \Psi_1\left|_{\eta=0} = q\right|_{\eta=0} = 0, \quad \left.
(\partial_\xi-\partial_\eta)^2\Psi_2\right|_{\eta=0} = -k^2e^{-ik\xi},$$
it follows that
$$ v_2 =-ik^2e^{-ik\xi} + iu_2(\xi,t)e^{-ik\xi} + \half \int^\eta_0d\eta'\bar q v_1.
\eqno (3.7b)$$
Equations (3.7) imply that the term $-ik^2\exp[-ik\xi]$ gives rise to $-ik^2\Psi$,
while the term
$$iu_2(\xi,t)e^{-ik\xi} = \inta dl\gamma(k-l,t)e^{-il\xi}, \eqno (3.8)$$
gives rise to the term $\inta dl\gamma(k-l,t)\Psi(l)$.  Also equation (3.8) implies
that $\gamma(k,t)$ is given by equation (3.3).

Equation (3.2) follows from the evaluation of equation (3.4a) at $\xi=0$. In this
respect we note:
$$\left. \Psi_{1_{\xi\eta}}\right|_{\xi=0} = -\half \left[ q_\eta\Psi_2 +
q\Psi_{2\eta} \right]_{\xi=0} =0,$$
since $q(0,\eta,t) = q_\eta(0,\eta,t)=0$.  Also
$$\left. \Psi_{1_{\xi\xi}}\right|_{\xi=0} =-\half \left[ q_\xi\Psi_2 +
q\Psi_{2\xi}\right]_{\xi=0} = -\half q_\xi(0,\eta,t),$$
since $\Psi_2(0,\eta,t,k)=1$.
\begin{flushright}
\textbf{QED}
\end{flushright}

\par
Using equation (3.2), it is now straightforward to obtain the time
evolution of $S(k,l,t)$:

\paragraph{Proposition 3.2}  Let $S(k,l,t)$ be defined by equation
(2.4), where $M^-_{22}(\xi,\eta,t,k)=\Psi_{2}\exp{[-ik\xi]}$,
$\Psi_{2}$ is defined in terms of $q(\xi,\eta,t)$ in equations (3.1),
and $q$ evolves in time according to equations (1.3)--(1.5).  Let
$$\hat S(\xi,\eta,t) = \inta \inta dkdle^{ik\xi + il\eta} S(k,l,t). \eqno (3.9)$$
Then $\hat S$ satisfies equations (1.8).

\paragraph{Proof.}  Using $(M^-_{12}, M^-_{22})^T = e^{-ik\xi} (\Psi_1,\Psi_2)^T$, it
follows that the vector $\Psi = (\Psi_1,\Psi_2)^T$ satisfies equations (3.1).  Also
$$
S(k,l,t) = \frac{1}{4\pi} \into \into d\xi d\eta q(\xi,\eta) e^{-il\eta}
\Psi_2(\xi,\eta,k) = \frac{1}{2\pi} \into d\eta e^{-il\eta}
\Psi_1(0,\eta,t,k), \eqno (3.10)
$$
where the first equality uses the definition of $S$ (see equation
(2.4)) and the second equality uses equation (3.1a).  Hence using
$\Psi_{1} (0,\eta,t,k) = \psi_{1}(\eta,t,k)$ and replacing in equation
(3.9) $S$ by the rhs of equation (3.10) we find
$$ 
\hat S(\xi,\eta,t) = \inta dke^{ik\xi}\psi_1(\eta,t,k). \eqno (3.11)
$$
In order to find the time evolution of $\hat S$ we need to take the $k$-Fourier
transform of equation (3.2).  In this respect we note that if
$q(0,\eta,t) = 0$, equations (3.1) imply
$$
\psi_1 = -\half \frac{q_\xi(0,\eta,t)}{k^2} + O\left( \frac{1}{k^3}\right), \quad
k\rightarrow \infty. \eqno (3.12)
$$
Using this estimate, equation (3.2) implies equation (1.8a).  In this
respect we note that the $k$-Fourier transform of $k^2\psi_{1} +
\frac{1}{2}q_{\xi}(0,\eta,t)$ equals $\hat{S}_{\xi\xi}$.  Indeed, if
$s(x)$ denotes the Fourier transform of $\psi$,
$$ 
s(x) = \inta e^{ikx} \psi(k)dk, \quad x>0, \eqno (3.13)
$$
and if $\psi(k)$ for large $k$ is given by
$$ 
\psi(k) = \frac{\alpha}{k^2} + O\left( \frac{1}{k^3}\right), \quad k \rightarrow
\infty, \eqno (3.14)
$$
then
$$ 
\frac{d^2s}{dx^2} = \inta e^{ikx} [\alpha -k^2\psi(k)]dk. \eqno (3.15)
$$
In order to derive this result we note that the definition of $s(x)$
can be rewritten as
$$ 
s(x) = \inta e^{ikx} \left[ \psi(k) + \frac{\alpha}{1+k^2}\right] dk + \alpha
\inta \frac{e^{ikx}}{1+k^2} dk;
$$
the second integral above can be computed explicitly and equals $\pi e^{-x}$.  Thus
$$
\frac{d^2s}{dx^2} = \inta e^{-ikx} \left[ \frac{\alpha k^2}{1+k^2} - \psi(k)\right]
dk + \alpha \pi e^{-x};
$$
simplifying the rhs of this equation we find (3.15).

Since $\psi_1(0,t,k) = \Psi_1(0,0,t,k) = 0$, equation (3.11) implies $\hat
S(\xi,0,t) = 0$.  Also
$$ \hat S(0,\eta,t) = \inta dk \psi_1(\eta,t,k) =0,$$
since $\psi_1$ is analytic for $\Im k <0$ and is of $O(\frac{1}{k^2})$ as $k\rightarrow
\infty$ (see equation (3.12)).
\begin{flushright}
\textbf{QED}
\end{flushright}

\section{Proof of Theorem 1.1}
In both sections 2 and 3 it was \emph{assumed} that $q(\xi,\eta,t)$
exists, it is smooth and has sufficient decay.  Then, under this
assumption, it was shown in section 2 that $q$ can be obtained through
the solution of a RH problem uniquely defined in terms of $S(k,l,t)$.
Furthermore, it was shown in section 3 that $S$ satisfies equations
(1.8).  We will now prove that $q(\xi,\eta,t)$ can be constructed through
the solution of the RH problem (1.9) \emph{without} the a priori
assumption of existence.  Furthermore, this result is valid
\emph{without} the need for the norms of
$q_0(\xi,t),u_1(\eta,t),u_2(\xi,t)$ to be small.

\par
The function $S_0(k,l)$ is defined through the solution
$(M_1(\xi,\eta),M_2(\xi,\eta))^T$ of the linear integral equations
(1.6).  These equations are of Volterra type, and thus if $q_0 \in L_1$, are
always solvable. 

\par
The Fourier transform $\hat{S}(\xi,\eta,t)$ of the function $S(k,l,t)$
satisfies the linear PDE (1.8a) with the initial condition $\hat{S}_0
(\xi,\eta)$ and with homogeneous Dirichlet boundary conditions.  Thus
if $u_1(\eta,t),u_2(\xi,t)$ have sufficient smoothness and decay,
$S(k,l,t)$ is well defined.

\par
If the functions $S,S_k,S_l \in L_2$ for fixed $t$, the RH problem (1.9) has
a global solution provided that its homogeneous version has only the
\emph{trivial} solution.  This is indeed the case [4]: Let $\mu$ satisfy
equations (1.9) but with $\mu=O(1/k)$ as $k \to \infty$.  The jump
conditions imply
$$
\begin{array}{l}
\mu_{11}^{+}(k) - \mu_{11}^{-}(k) = -\inta dl
\overline{S(l,k)} e^{-il\xi - ik\eta} \mu_{12}^{+} (l), \\
\mu_{12}^{+}(k) - \mu_{12}^{-}(k) = -\inta dl
S(k,l) e^{il\eta + ik\xi} \mu_{11}^{-} (l),
\end{array} \eqno{(4.1)}
$$
where for convenience of notation we have suppressed the $\xi,\eta,t$
dependence of $\mu,S$.  Multiplying equation (4.1a) by
$\overline{\mu_{11}^{-}(\bar{k})}$, equation (4.1b) by
$\overline{\mu_{22}^{+}(\bar{k})}$, integrating over $k$, and adding
the resulting equations we find an equation whose lhs equals
$$
\inta dk \left( |\mu_{11}^{-}(k)|^2 + |\mu_{12}^{+}(k)|^2 \right),
$$
while the rhs is imaginary (in this derivation  we have used that the
integrals of $\overline{\mu_{11}^{-}(\bar{k})}\mu_{11}^{+}(k)$ and of
$\overline{\mu_{22}^{+}(\bar{k})}\mu_{12}^{-}(k)$ vanish since these
products are analytic for $\Im k>0$ and $\Im k<0$ respectively, and
are of $O(1/k^2)$ as $k \to \infty$).  Hence, $\mu_{11}^{-} =
\mu_{12}^{+} = 0$.

\par
For the proof of the fact that $q(\xi,\eta,0)=q_0(\xi,\eta)$ we note that
$q(\xi,\eta,0)$ is defined through the RH problem (1.9) but with
$S_0(k,l)$ instead of $S(k,l,0)$.  By repeating the analysis of section 2 where
$q(\xi,\eta,t)$ is now replaced by the given function $q_0(\xi,\eta)$, it follows
that $q_0(\xi,\eta)$ is characterised by precisely the same RH problem
as the problem characterising $q(\xi,\eta,0)$. (We note that $(M_1,M_2)^T
=(M_{12}^{-},M_{22}^{-})^T$).  Hence the unique solution of this RH problem implies
$q(\xi,\eta,0)=q_0(\xi,\eta)$.  

\par
We now show that if $q(\xi,\eta,t)$ is defined through the solution of
the RH problem (1.9) then $q$ satisfies the DSI equation.  This proof is based
on the extended version of the dressing method presented in [14].  The
dressing method can be used to show that if $M$ solves an appropriate RH
problem then $M$ also solves both parts of the associated Lax pair.
Hence, using the compatibility condition of this pair, $q$ solves the
relevant nonlinear PDE. 

\par
We first show that $M$ solves the $t$-independent part of the Lax pair
$$
M_x + \sigma_3 M_y - ik[\sigma_3,M] +QM = 0, \eqno{(4.2)}
$$
where
$$ 
\xi = x+y, \quad \eta = x-y, \quad \sigma_3 = \mathrm{diag}(1,-1), \quad Q =
\left(
\begin{array}{cc}
0 & q \\
-\bar{q} & 0
\end{array} \right), \eqno{(4.3)}
$$
and $[,]$ denotes the usual matrix commutator.  Indeed, the jump
condition of the RH problem can be written in the form
$$
M^{+}(x,y,t,k) - M^{-}(x,y,t,k) = \inta dl \tilde{M}(x,y,t,l)
F(x,y,t,k,l), \eqno{(4.4)}
$$
where $\tilde{M}$ denotes the matrix with first, second column vectors
$(M_{11}^{-},M_{21}^{-})^T$, $(M_{12}^{+},M_{22}^{+})^T$ respectively,
and $F$ denotes the off-diagonal matrix with
$$
\begin{array}{l}
F_{12}(x,y,t,k,l) = -S(k,l,t) e^{ik(x+y) + il(x-y)}, \\
F_{21}(x,y,t,k,l) = \overline{F_{12}(x,y,t,l,k)}.
\end{array} \eqno{(4.5)}
$$
Using
$$
\begin{array}{l}
{F_{12}}_{x} = i(k+l)F_{12}, \quad {F_{21}}_{x} = -i(k+l)F_{21}, \\
{F_{12}}_{y} = i(k-l)F_{12}, \quad {F_{21}}_{y} = i(k-l)F_{21},
\end{array}
$$
and writing these equations in a matrix form we find
$$
F_{x} = il\sigma_{3}F -ikF\sigma_{3}, \quad F_{y} =
i(k-l)F. \eqno{(4.6)}
$$
Let the operators $D_{x}, D_{y}$ be defined by
$$
D_{x}M = M_{x} + ikM\sigma_{3}, \quad D_{y}M = M_{y} -
ikM. \eqno{(4.7)}
$$
These definitions and equations (4.6) imply that
$$
D_{x} \inta dl \tilde{M}(l)F(k,l) = \inta dl \left(D_{x}\tilde{M}(l)
\right)F(k,l),
$$
$$
D_{y} \inta dl \tilde{M}(l)F(k,l) = \inta dl \left(D_{y}\tilde{M}(l)
\right)F(k,l),
$$
where for simplicity of notation we have suppressed the $x,y,t$
dependence.  Hence, both $D_{x}M$ and $D_{y}M$ satisfy the same jump
condition as $M$.  Using the fact that $QM$ also satisfies the same
jump condition, it follows that the expression
$$
D_{x}M + \sigma_{3}D_{y}M + QM \eqno{(4.8)}
$$
satisfies the same jump condition as $M$.  Using
$$
M = I + \frac{M^{(1)}(x,y,t)}{k} + \frac{M^{(2)}(x,y,t)}{k^2} +
O\left(\frac{1}{k^3}\right), \eqno{(4.9)}
$$
it follows that the large $k$ behaviour of the expression (4.8) is
given by
$$
\left( Q - i[\sigma_{3}, M^{(1)}] \right) + O\left(\frac{1}{k}\right).
$$
Hence if we define $Q$ by
$$
Q(x,y,t) = i[\sigma_{3}, M^{(1)}], \eqno{(4.10)}
$$
the expression (4.8) satisfies the homogeneous version of the RH
problem, therefore it vanishes and $M$ satisfies equation (4.2).  We
note that $Q$ is off diagonal; furthermore equation (4.5b) implies
certain symmetry relations for $M$, which in turn implies that
$(Q)_{21} = -\overline{(Q)}_{12}$.  In summary, if $M$ satisfies the
RH problem (1.9) and if $q$ is defined through $M$ by equation (1.11), then
$M$ satisfies the $t$-independent part of the Lax pair.

\par
We now show that $M$ also satisfies the $t$-part of the Lax pair.  For
this purpose, rather than assuming that $S$ satisfies equation (1.8a)
we first assume that $S(k,l,t)$ satisfies the equation
$$
iS_{t} - (k^2+l^2)S + \inta d\nu (-i)\gamma_{1}(\nu-l,t) S(k,\nu,t) +
i\gamma_{2}(\nu,t)S(k-\nu,l,t), \eqno{(4.11)}
$$
where the functions $\gamma_{1},\gamma_{2}$ are defined by
$$
\gamma_{1}(\nu,t) = \frac{i}{2\pi} \inta d\eta u_{1}(\eta,t)e^{i\nu
  \eta}, \quad \gamma_{2}(\nu,t) = -\frac{i}{2\pi} \inta d\xi
  u_{2}(\xi,t)e^{-i\nu \xi}. \eqno{(4.12)}
$$
These equations imply that $F$ satisfies the evolution equation
$$
F_{t} = ik^2 F\sigma_{3} - il^2 \sigma_{3} F + \inta d\nu \Big(
\Gamma(x,y,t,\nu-l) F(x,y,t,k,\nu) - F(x,y,t,k-\nu,l)
\Gamma(x,y,t,\nu) \Big), \eqno{(4.13)}
$$
where $\Gamma$ is defined by
$$
\Gamma(x,y,t,\nu) = e^{i\nu y} {\mathrm{diag}} \left(e^{-i\nu
  x}\gamma_{1}(\nu), e^{i\nu x}\gamma_{2}(\nu) \right). \eqno{(4.14)}
$$
Indeed, using 
$$
F_{12}(x,y,t,k,l) = -S(k,l,t)e^{ik\xi + il\eta}, \;
\Gamma_{11}(x,y,t,\nu) = e^{-i\nu\eta}\gamma_{1}(\nu,t), \;
\Gamma_{22}(x,y,t,\nu) = e^{i\nu\xi}\gamma_{2}(\nu,t),
$$
the $(12)$ component of equation (4.13) yields equation (4.11).

\par
The time dependence of $F$ suggests the introduction of the operator
$D_{t}$ defined by
$$
D_{t}M = M_{t} -ik^2 M \sigma_{3} + \inta d\nu M(x,y,t,k-\nu)
\Gamma(x,y,t,\nu). \eqno{(4.15)}
$$
Indeed it can be shown that
$$
D_{t} \inta dl \tilde{M}(l)F(k,l) = \inta dl
\big(D_{t}\tilde{M}(l)\big) F(k,l).
$$ 
This equation is a direct consequence of the definition of $D_{t}$ and
of equation (4.13) provided that
$$
\inta dl \inta d\nu \tilde{M}(l) \Gamma(\nu-l) F(k,\nu) = \inta dl
\inta d\nu \tilde{M}(l-\nu) \Gamma(\nu) F(k,l);
$$
this equality is established by replacing in the rhs $l-\nu$ by $\nu$
and then exchanging $l$ and $\nu$ in the resulting equation.

\par
Since $D_{t}M$ satisfies the same jump condition as $M$, following
the same logic as in the derivation of equation (4.2) it follows that
$$
-D_{t}M + i\sigma_{3}{D_{y}}^{2}M + BD_{y}M + AM = 0, \eqno{(4.16)}
$$
where the matrices $A(x,y,t)$ and $B(x,y,t)$ will be chosen by the
requirement that the lhs of equation (4.16) is of $O(1/k)$ as $k
\to \infty$.  In this respect we first introduce the following
notations for the $O(1/k)$ term of $M$, see equation (4.9),
$$
M^{(1)}(x,y,t) =
\left(
\begin{array}{cc}
a & b \\
c & d
\end{array} \right). \eqno{(4.17)}
$$

\par
Equation (4.10) yields
$$
b = \frac{q}{2i}, \quad c = \frac{\bar{q}}{2i}. \eqno{(4.18)}
$$
The $O(1/k)$ term of equation (4.2) implies
$$
M^{(1)}_{x} + \sigma_{3}M^{(1)}_{y} - i[\sigma_{3},M^{(2)}] + QM^{(1)}
= 0. \eqno{(4.19)}
$$
The diagonal and off-diagonal parts of this equation yield
$$
\partial{}_{x}M^{(1)}_{D} + \sigma_{3}\partial{}_{y}M^{(1)}_{D} + QM^{(1)}_{O} = 0,
\eqno{(4.20)}
$$
$$
\partial{}_{x}M^{(1)}_{O} + \sigma_{3}\partial{}_{y}M^{(1)}_{O} -
i[\sigma_{3},M^{(2)}_{O}] + QM^{(1)}_{D} = 0. \eqno{(4.21)}
$$
Equation (4.20) implies
$$
a_{\xi} = -\frac{|q|^2}{4i}, \quad d_{\eta} =
\frac{|q|^2}{4i}. \eqno{(4.22)}
$$

\par
Using the definitions of $D_{t}$ and $D_{y}$, equation (4.16) becomes
$$
M_{t} = -ik^2[\sigma_{3},M] - ikBM + 2k\sigma_{3}M_{y} + AM + BM_{y} +
i\sigma_{3}M_{yy}
$$
$$
\hspace{5cm} - \inta d\nu
M(x,y,t,k-\nu)\Gamma(x,y,t,\nu). \eqno{(4.23)}
$$
The $O(k)$ term of this equation yields
$$
B = -[\sigma_{3},M^{(1)}],
$$
thus (comparing with equation (4.10))
$$
B = iQ, \eqno{(4.24)}
$$
The $O(1)$ term of equation (4.23) yields
$$
-i[\sigma_{3},M^{(2)}] - iBM^{(1)} + 2\sigma_{3}M^{(1)}_{y} + A -
\inta d\nu \Gamma(x,y,t,\nu) = 0. \eqno{(4.25)}
$$
Using equation (4.19) to replace the first two terms of this equation
we find
$$
-M^{(1)}_{x} + \sigma_{3}M^{(1)}_{y} + A - \inta d\nu
\Gamma(x,y,t,\nu) = 0.
$$
Solving this equation for $A$ and using the definition of $\Gamma$
(equations (4.12)) we find
$$
A_{11} = 2a_{\eta} + iu_{1}, \quad A_{12} = 2b_{\eta}, \quad A_{21} =
2c_{\xi}, \quad A_{22} = 2d_{\xi} - iu_{2}. \eqno{(4.26)}
$$
In summary, if the time evolution of $S(k,l,t)$ is given by equation
(4.11), where $\gamma_{1}$ and $\gamma_{2}$ are defined by equations
(4.12), then the time evolution of $M$ is given by (4.23) where $B=iQ$
and the components of the matrix $A$ are given by equations (4.26).

\par
Letting
$$
M = \Psi \mathrm{diag} \left(e^{-ik\eta},e^{ik\xi}\right),
\eqno{(4.27)}
$$
and using that the $y$-derivative of the above diagonal matrix equals
$ik$ times this matrix, it follows that the time evolution of $\Psi$
is given by
$$
\Psi_{t} = i\sigma_{3}\partial{}^{2}_{y} \Psi + iQ\partial{}_{y} \Psi
+ A \Psi + ik^2\Psi\sigma_{3} - \inta d\nu \Psi(x,y,t,k-\nu)
\mathrm{diag}\Big(\gamma_{1}(\nu,t),\gamma_{2}(\nu,t)\Big). \eqno{(4.28)}
$$
With a particular choice of the constants of integration, equations
(4.22) yield
$$
a = \frac{i}{4} \int_{0}^{\xi} d\xi' |q|^2, \quad d = -\frac{i}{4}
\int_{0}^{\eta} d\eta' |q|^2. \eqno{(4.29)}
$$
Then the first and the fourth of equations (4.26) become
$$
A_{11} = i\left(\frac{1}{2}\int_{0}^{\xi} d\xi' |q|^{2}_{\eta} +
u_{1}(\eta,t) \right), \quad A_{22} =
-i\left(\frac{1}{2}\int_{0}^{\eta} d\eta' |q|^{2}_{\xi} + u_{2}(\xi,t)
\right).
$$
Using these expressions as well as equations (4.18), equations (4.26)
become
$$
A_{11} = iU_{1}, \quad A_{12} = -iq_{\eta}, \quad A_{21} =
-i\bar{q}_{\xi}, \quad A_{22} = -iU_{2}. \eqno{(4.30)}
$$
Denoting by $(\Psi_{1},\Psi_{2})^{T}$ the second column vector of
$\Psi$, replacing $A$ in equation (4.28) by equations (4.30), and
noting that $\gamma_{2}=-\gamma$, where $\gamma$ is defined by
equation (3.3), equation (4.28) implies that $\Psi_{1},\Psi_{2}$
satisfy precisely equations (3.4) where the vector $v$ is defined by
equation (3.5).

\par
The vector $(\Psi_{1},\Psi_{2})^{T}$ also satisfies
$$
{\Psi_{1}}_{\xi} = -\frac{1}{2}q\Psi_{2}, \quad {\Psi_{2}}_{\eta} =
\frac{1}{2} \bar{q}\Psi_{1}. \eqno{(4.31)}
$$
It can be verified that the compatibility of equations (3.4) and
(4.31) yields equations (1.3).

\par
It remains to: (a) show that the Fourier transform of equation (4.11)
yields equation (1.8a); (b) justify the choice of the constants of
integration of equations (4.22); (c) establish that
$q(0,\eta,t)=q(\xi,0,t)=0$.  These interelated facts can be proved by
``reversing'' the relevant arguments used in section 3.

\begin{flushright}
\textbf{QED}
\end{flushright}

\section{The $t$-part of the Lax pair for the non-homogeneous case.}

For simplicity of notation throughout this section  we suppress the
$t$-dependence.
\paragraph{Proposition 5.1}  Let the vector $\Psi$ satisfy equations (3.1) where
$q$ satisfies equations (1.3) on the quarter plane.  Then the function
$\psi_1(\eta,k) = \Psi_1(0,\eta,k)$ satisfies the following equation:
$$ 
i\psi_{1_t} + \psi_{1_{\eta\eta}} - \left[ k^2\psi_1 + \frac{ikq}{2} +
\frac{q_\xi}{2} + \frac{q}{8} \int^\eta_0 d\eta'|q|^2\right] +
u_{1}\psi_1
$$
$$ 
-i\inta dl \left[ \gamma(k-l) + \delta(k-l,k) + \varepsilon(k-l,k)\right]
\psi_1(\eta,l)
$$
$$ 
- \frac{q_\xi}{4} \int^\eta_0 d\eta' \bar q\psi_1 + \frac{q}{4} \int^\eta_0 d\eta'
\bar q_\xi \psi_1 - \frac{1}{16} \int^\eta_0 d\eta'|q|^2 \int^{\eta'}_0 d\hat \eta
\bar q\psi_1 = 0, \eqno (5.1)
$$
where: $q, q_\xi$ in the first bracket are evaluated at $(0,\eta)$;
$|q|, \bar{q}\psi_{1}, \bar{q}_{\xi}\psi_{1}$ in the integrals with
respect to $d\eta'$ are evaluated at $(0,\eta')$; $\bar{q}\psi_{1}$ in
the integral with respect to $d\hat{\eta}$ is evaluated at
$(0,\hat{\eta})$; $\gamma$ is defined by equation (3.3); and $\delta,
\varepsilon$ are defined by
$$ 
\delta(l,k) = \frac{i}{8\pi} \into \into d\xi d\hat \xi \left[ \bar q_\eta(\xi,0)
q(\hat \xi + \xi,0) - \bar q(\xi,0)q_\eta(\hat \xi + \xi,0)\right]e^{-ik\hat\xi
-il\xi}, \eqno (5.2)
$$
$$
\varepsilon(l,k) = - \frac{i}{32\pi}\into \into \into d\xi d\hat \xi d\xi' \bar
q(\xi,0) |q(\hat \xi+\xi,0|^2q(\xi+\hat\xi+\xi',0)e^{-ik\hat\xi - ik\xi'-il\xi}. \eqno
(5.3)
$$
\paragraph{Proof}  We first show that $\Psi$ satisfies equations (3.4) where $v$ is
defined by equation (3.5) with the additional term
$$\inta dl \left(\delta (k-l,k) + \varepsilon(k-l,k)\right) \Psi(\xi,\eta,l).$$
Indeed, proceeding as in the proof of Proposition 3.1 we find that the forcing of
the equation satisfied by $v_2$ involves the following additional terms
$$ \frac{i}{4} \left[ \bar q_\eta(\xi,0) \int^\infty_\xi d\xi' q(\xi',0) e^{-ik\xi'} -
\bar q(\xi,0) \int^\infty_\xi d\xi' q_\eta(\xi',0) e^{-ik\xi'}\right] $$
$$ - \frac{i}{16} \bar q(\xi,0) \int^\infty_\xi d\xi' |q(\xi',0)|^2
\int^\infty_{\xi'} d\xi'' q(\xi'',0) e^{-ik\xi''}. \eqno (5.4)$$
For the computation of these additional terms we have used the
following expressions, where for convenience of notation we have
suppressed the $k$-dependence:
$$\begin{array}{ll} \psi_1(\xi,0) = \half \int^\infty_\xi d\xi'q(\xi',0) e^{-ik\xi'}, \quad
\psi_2(\xi,0) = e^{-ik\xi}, \\ \\
\psi_{1_\xi}(\xi,0) = -\half q(\xi,0) e^{-ik\xi},\\ \\
\psi_{1_\eta}(\xi,0) = \half \int^\infty_\xi d\xi' q_\eta(\xi',0) e^{-ik\xi'} +
\frac{1}{4} \int^\infty_\xi d\xi'|q(\xi',0)|^2\psi_1(\xi',0),\\ \\
\psi_{2_{\xi\xi}}(\xi,0) =-k^2e^{-ik\xi},\\ \\
\psi_{2_{\eta\xi}}(\xi,0) = \half \bar q_\xi(0,\eta) \psi_1(\xi,0) -
\frac{1}{4}|q|^2(\xi,0) e^{-ik\xi},\\ \\
\psi_{2_{\eta\eta}}(\xi,0) = \half \bar q_\eta(\xi,0)\psi_1(\xi,0) + \half \bar
q(\xi,0) \psi_{1_\eta} (\xi,0). \end{array} \eqno (5.5)$$
Denoting the first bracket in (5.4) by
$$ \int^\infty_\xi d\xi' A(\xi,\xi') e^{-ik\xi'} = \inta dl\delta(k-l,k)e^{-il\xi},$$
replacing $k-l$ by $l$ and $\xi'-\xi$ by $\hat \xi$, we find
$$\into d\hat\xi A(\xi,\hat \xi+\xi) e^{-ik\hat \xi} = \inta dl\delta
(l,k)e^{il\xi},$$
which yields (5.2).  Similarly denoting the second term in (5.4) by
$$ \int^\infty_\xi d\xi' A(\xi,\xi') \int^\infty_{\xi'} d\xi''B(\xi'') e^{-ik\xi''} =
\inta dl\varepsilon (k-l,k)e^{-il\xi},$$
replacing $k-l$ by $l$, $\xi''-\xi'$ by $\hat{\hat \xi}$, and $\xi'-\xi$ by $\hat
\xi$, we find
$$ \into d\hat \xi A(\xi,\hat \xi + \xi) \into d\hat{\hat \xi} B(\hat{\hat \xi} + \hat
\xi + \xi) e^{-ik(\hat\xi + \hat{\hat \xi})} = \inta dl \varepsilon (l,k)e^{il\xi},$$
which yields (5.3).

Equation (5.1) follows by evaluating equation (3.4a) at $\xi=0$.  For this computation
we use the following equations, where again for convenience of
notation we have suppressed the $k$-dependence:
$$\begin{array}{ll}
\psi_2(0,\eta) = 1 + \half \int^\eta_0 d\eta' \bar q(0,\eta')\psi_1(0,\eta'), \quad
\psi_{2_\eta}(0,\eta) = \half \bar q(0,\eta)\psi_1(0,\eta), \\ \\ 
\psi_{2_\xi}(0,\eta)=-ik + \half \int^\eta_0 d\eta' \bar q_\xi (0,\eta') \psi_1(0,\eta') -
\frac{1}{4} \int^\eta_0 d\eta' |\bar q(0,\eta'|^2\psi_2(0,\eta'),\\ \\
\psi_{1_{\xi\xi}}(0,\eta) =- \half q_\xi (0,\eta)\psi_2(0,\eta) -  \half
q(0,\eta)\psi_{2_\xi}(0,\eta),\\ \\
\psi_{1_{\xi\eta}}(0,\eta) =- \half q_\eta(0,\eta)\psi_2(0,\eta) - \frac{1}{4}
|q(0,\eta)|^2\psi_1(0,\eta).\end{array}$$
\begin{flushright}
\textbf{QED}
\end{flushright}

\paragraph{Proposition 5.2}  Let $S(k,l,t)$ be defined by equation (2.4), where
$M^-_{22}=\Psi_{2}\exp{[-ik\xi]}$, $\Psi$ is defined in terms of $q$
in equations (3.1) and $q$ evolves in time according to equation (1.3).  Let
$\hat S$ be defined in terms of $S$ by equation (3.9).  Then $\hat S$
satisfies the equation
$$i\hat S_t + \hat S_{\xi\xi} + \hat S_{\eta\eta} + (u_1+u_2) \hat S + \int^\xi_0
d\tilde\xi F_1(\xi,\tilde \xi,t) \hat S(\tilde \xi,\eta,t) + \int^\eta_0 d\tilde \eta
F_2(\eta, \tilde\eta,t) \hat S(\xi,\tilde\eta,t) = 0, \eqno (5.7)$$
where
$$ F_1(\xi,\tilde\xi,t) = \frac{1}{4} \left[ \bar q_\eta(\tilde\xi,0) q(\xi,0) - \bar
q(\tilde \xi,0)q_\eta(\xi,0)\right] - \frac{1}{16} q(\xi,0) \bar q(\tilde \xi,0)
\int^\xi_{\tilde \xi}d\xi'|q(\xi',0)|^2,$$
$$F_2(\eta,\tilde\eta,t) = \frac{1}{4} \left[ \bar q_\xi(0,\tilde\eta)q(0,\eta) - \bar
q(0,\tilde\eta)q_\xi(0,\eta)\right] - \frac{1}{16} q(0,\eta)\bar q(0,\tilde\eta)
\int^\eta_{\tilde \eta} d\eta'|	q(0,\eta')|^2. \eqno (5.8)$$
Furthermore, $\hat S$ satisfies the boundary conditions
$$ \hat S(\xi,0,t) = \pi q(\xi,0), \quad \hat S(0,\eta,t) = \pi q(0,\eta). \eqno
(5.9)$$
\paragraph{Proof}  The analogue of equation (3.12) is now
$$ \psi_1 =   \frac{ - \frac{iq(0,\eta)}{2}}{k} - \frac{ \frac{q_\xi}{2} + \frac{q}{8}
\int^\eta_0 d\eta'|q(0,\eta')|^2}{k^2} + O \left( \frac{1}{k^3}\right), \quad k
\rightarrow \infty. \eqno (5.10)$$
Equation (3.11) yields
$$\hat S(\xi,0,t) = \inta dke^{ik\xi} \Psi_1(0,0,t,k) = \half \inta dke^{ik\xi}
\into d\xi' q(\xi',0) e^{-ik\xi'} = \pi q(\xi,0).$$
Furthermore,
$$ \hat S(0,\eta,t) = \inta dk\Psi_1(0,\eta,t,k) = \inta dk  \left[
  \Psi_1(0,\eta,t,k) + \frac{i}{2} \frac{q(0,\eta)}{k} \right] -
  \frac{i}{2} \inta \frac{dk}{k} q(0,\eta), = \pi q(0,\eta),$$
since $\Psi_1(0,\eta,t,k)$ is analytic for $\Im k <0$.

Taking the Fourier transform of equation (5.1) and using the estimate (5.10) we find
equation (5.7).  In this respect we note that if we denote the bracket in the rhs of
equation (5.2) by $A(\xi,\hat \xi)$, then the contribution of the term involving
$\gamma$ is given by
$$-i\inta dl \inta dk\gamma(k-l,t)\psi_1(l) e^{ik\xi} = \frac{1}{8\pi} \inta dl \inta
dk \into d\xi' \into d\hat\xi A(\xi',\hat\xi) e^{-ik\hat\xi -i(k-l)\xi'+ik\xi}
\psi_1(l)$$
$$ = \frac{1}{4} \into d\xi' \into d\hat\xi A(\xi',\hat\xi) \delta (\hat
\xi-(\xi-\xi')) \hat S(\xi',\eta,t) = \frac{1}{4}
\into d\xi'\theta(\xi-\xi')A(\xi',\xi-\xi')\hat S(\xi',\eta,t),$$
where $\psi_{1}(l) = \psi_{1}(\eta,l)$.  Similarly the contribution of
the term involving $\delta$ is given by
$$-\frac{1}{32} \inta \inta dkdl \into\into\into d\tilde \xi d\hat \xi d\xi' \bar
q(\tilde\xi,0) |q(\hat\xi+\tilde\xi,0|^2q(\tilde\xi + \hat\xi + \xi',0) e^{-ik\hat\xi
-ik\xi' -i(k-l)\tilde\xi + ik\xi}\psi_1(l)$$
$$=-\frac{q(\xi,0)}{16} \into d\tilde \xi \into d\hat\xi \theta(\xi-\tilde\xi-\hat\xi)
\bar q(\tilde\xi,0) |q(\hat\xi+\tilde\xi,0)|^2\hat S(\tilde \xi,\eta)$$
$$ = -\frac{q(\xi,0)}{16} \int^\xi_0 d\tilde \xi \int^{\xi-\tilde\xi}_0 d\hat \xi \bar
q(\tilde \xi,0) |q(\hat\xi + \tilde \xi,0)|^2\hat S(\tilde \xi,\eta).$$
The area of integration is depicted in Figure 5.1a.  Making the change of
\begin{figure}[h]
\begin{center}
\begin{minipage}[b]{6cm}
 \centerline
  {\epsfbox{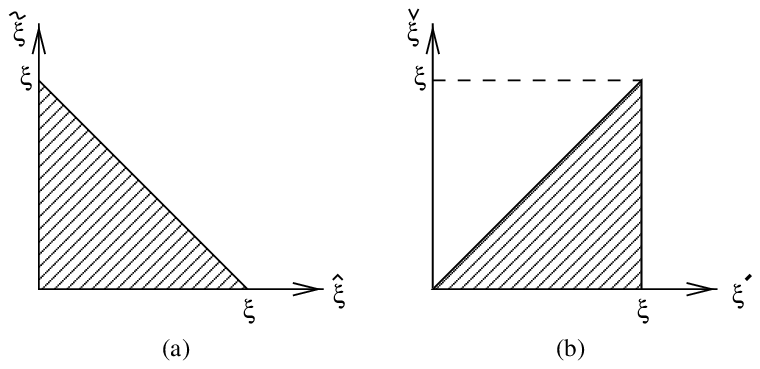}}
\centerline{\bf Figure 5.1}
\end{minipage}
\end{center}
\end{figure}
variables $\xi'=\hat\xi + \tilde\xi$, $\check\xi = \tilde\xi$, the area of
integration is mapped to the area depicted in Figure 5.1b. Thus the relevant integral
becomes
$$ \frac{-q(\xi,0)}{16} \int^\xi_0 d\check\xi \left( \int^\xi_{\check \xi}d\xi'
|q(\xi',0)|^2\right) \bar q(\check\xi,0) \hat S(\check\xi,\eta).$$
Also regarding the last term of equation (5.1) we note that taking its Fourier
transform we find
$$ \frac{-q(0,\eta)}{16} \int^\eta_0 d\eta' |q(0,\eta')|^2 \int^{\eta'}_0 d\hat\eta
\bar q(0,\hat \eta) \hat S(\xi,\hat\eta,t).$$
Changing the order of the integration we find that the relevant contribution involves
the last term of $F_2$, see equation (5.8).
\begin{flushright}
\textbf{QED}
\end{flushright}

\section{Discussion}
In the last decade considerable progress has been made in the
understanding of boundary-value problems for integrable nonlinear
evolution PDEs in \emph{one} spatial dimension. For example, two of
the articles in this special issue are concerned with this
development [9],[10]. Here, to our knowledge for the first time, a
boundary-value problem is solved in \emph{two} spatial dimensions. 

It has been emphasised by the author that an important difference
between initial- and boundary-value problems is the following: For
initial-value problems one needs to perform the spectral analysis
of the $t$-independent part of the Lax pair only, while for boundary-value
problems the spectral analysis of \emph{both} parts of the Lax pair is
needed. In this respect we recall the following developments: 
\begin{enumerate}
\item[(i)] It is interesting that the spectral analysis of the
  $t$-part of the Lax pair (in addition to that of the $t$-independent
  part), was first considered not for an equation in one spatial
  dimension but for an equation in two spatial dimensions, namely the
  DSI equation: In [4] equations (1.3) were solved on the plane, but
  with nontrivial boundary conditions at infinity, 
$$
U_{1}(\xi,\eta,t) \to u_{1}(\eta,t), \; \xi \to \infty; \quad
U_{2}(\xi,\eta,t) \to u_{2}(\xi,t), \; \eta \to \infty.
$$
The authors of [4], rather than performing the explicit spectral
analysis of the relevant $t$-dependent eigenvalue equation they made
use of a certain completeness relation (of course the derivation of
this relation is based on the spectral analysis of the \s eigenvalue
equation).

\item[(ii)] In [11] the \emph{independent} spectral analysis of the
  two parts of the Lax pair of the nonlinear \s equation led to the
  formulation of the solution in terms of \emph{two} Riemann-Hilbert
  (RH) problems, which had to be solved in sequence. 

\item[(iii)] In [12] the above two RH problems were combined and the
  solution $q$ was expressed in terms of a \emph{single} RH
  problem. This RH problem has the distinctive and very useful
  feature of involving jump matrices with \emph{explicit} exponential
  $(x,t)$ dependence.

\item[(iv)] It was shown in [13] that the above RH problem can be
  derived in a straightforward manner by performing the
  \emph{simultaneous} spectral analysis of the Lax pair.  The rigorous
  proof that the solution of this RH problem yields the unique solution of
  the given initial-boundary value problem was presented in [7],[8]. 
\end{enumerate}

\par
In the present article we have implemented for equations (1.3)--(1.5)
the construction of (i) above.  The main reason for using (i) instead
of (iv) is the fact that for the case of homogeneous Dirichlet
boundary conditions the analysis of the $t$-part of the Lax pair is
very similar to the analysis presented in [4].

\par
The implementation of (iv) for the initial-boundary value problem
formulated in Proposition 1.1 remains open.  Furthermore, the analysis
of the associated global relation, which characterises $g_1$ and $f_1$
in terms of $g_0,f_0$ and $q_0$, also remains open.

\end{document}